\documentstyle[11pt]{article}

\font\mbld=cmmib10 scaled \magstep1
\def\ttt{\mbld \char'034}

\begin{document}

\date{}

\begin{titlepage}

\title{Properties of  predictive formulation of
the Nambu-Jona-Lasinio  model and ghost problem}

\author{
 V.E. Rochev\\{\it Institute for High Energy Physics, 142280,
Protvino, Moscow region,  Russia} }
\date{}
\date{}

\end{titlepage}
\maketitle
\begin{abstract}
Recently proposed  by Battistel et al.  "predictive formulation of
the NJL model" \quad is discussed and its connection with the
differential regularization is noted. The principal problem of
this formulation is a non-physical singularity (Landau pole) in
meson propagators.  A modification of the formulation, which is
free of the Landau pole and conserves main features of the
approach, is proposed.

\end{abstract}


\newpage

Effective models  in the non-perturbative region are a
considerable part of the modern strong-interaction theory. One of
the most successful effective models of strong interaction of
light hadrons is the Nambu-Jona-Lasinio (NJL) model with quark
content. Since the NJl model in the leading approximation includes
the quark loops and is based on non-renormalized four-fermion
interaction, the essential aspect of this model is a
regularization, which on the widely current opinion constitutes a
part of the definition of the NJL model. Though predictions of the
NJL model for commonly used regularizations (such as the cutoff in
momentum space, the Pauli-Villars regularization or the
proper-time regularization) are very similar (see, e.g.,
\cite{kle} for review), nevertheless such dependence on the
regularization prescription cannot be satisfactory from the
general theoretic point of view.

A very interesting attempt was made recently in the  work of
Battistel {\it et al} \cite{BDK} for releasing the NJL model from
the regularization dependence. In this work a  method for the
definition of one-loop Green functions of the NJL model was
proposed (see also foregoing works \cite{BatNem}). The idea of the
method is to avoid the explicit evaluation of divergent integrals
with any specific regularization. The finite parts are separated
of the divergent ones and are integrated without regularization.
Then the NJL model becomes predictive in the sense that its
consequences do not depend on the specific regularization of
divergent integrals. An important result of work \cite{BDK} is a
proof of the fulfilment of all symmetry constraints on
leading-order Green functions. A
 parameter choice  leads to the
reasonable values of the constituent quark mass and other model
parameters.

In the present paper some features of the calculational scheme of
work \cite{BDK}\footnote{We shall name the calculational scheme of
work \cite{BDK} as the BDK approach, or the implicit
regularization.} are analyzed,  and a connection of the BDK
approach with the
 differential regularization  is
pointed.  The principal problem of the BDK approach
 is  the singularity of meson propagators in the Euclidean momentum
 region. The presence of this singularity (Landau pole)
prevents  meson-loop calculations and, therefore, makes impossible
any calculations of corrections to the leading approximation. A
modification of the scheme, which is free of Landau pole and
conserves the main features of the approach of \cite{BDK}, is
proposed.

Following \cite{BDK} we consider the simplest physically
non-trivial variant of the NJL model with chiral symmetry
$SU_V(2)\times SU_A(2)$. The model Lagrangian is
\begin{equation}
{\cal L}=\bar \psi (i\hat \partial-m_0)\psi+\frac{g}{2}
\biggl[(\bar\psi\psi)^2+(\bar\psi i\gamma_5
\mbox{\ttt}\psi)^2\biggr]. \label{LSU2}
\end{equation}
Here $ \psi$ is the quark field with  $n_c$ colours, $m_0$ is the
current quark mass, $g$ is the coupling constant of  $m^{-2}$
dimension and  $\mbox{\ttt}$ are Pauli matrices.

The leading approximation of the model is the mean-field
approximation, which coincides in the case  with the leading order
of $1/n_c$--expansion. All Green functions of the leading
approximation are expressed in terms of quark one-loop integrals.
A problem of calculations of these integrals is reduced to the
definition of following five divergent integrals (see \cite{BDK}):

\begin{equation}
\{I_1;\;I_1^\mu\}=\int \frac{d^4k}{(2\pi)^4}
\frac{\{1;\;k^\mu\}}{(k+k_1)^2-M^2} \label{I1}
\end{equation}
\begin{equation}
\{I_2;\;I_2^\mu;\;I_2^{\mu\nu}\}=\int \frac{d^4k}{(2\pi)^4}
\frac{\{1;\;k^\mu;\;k^\mu k^\nu\}}{[(k+k_1)^2-M^2][(k+k_2)^2-M^2]}
\label{I2}
\end{equation}
Here $M$ is the dynamical (constituent) quark mass, which is a
non-trivial solution of the gap equation for the NJL model.

A basic device for the definition of integrals (\ref{I1}) and
(\ref{I2}) in the BDK approach is an algebraic identity for the
propagator function (see equation (28) in work \cite{BDK}). With
the identity the divergent parts of integrals (\ref{I1}) and
(\ref{I2}) have been expressed via three tensorial and two scalar
external-momentum-independent integrals which were treated in the
sense of some unspecified regularization. Then integrals
(\ref{I1}) and (\ref{I2}) have been expressed in terms of these
five integrals and two standard convergent integrals  (see
formulae of section III in work \cite{BDK}).

In connection with these expressions  we draw attention to some
general regularization-independent property of integrals
(\ref{I1}) and (\ref{I2}). Namely, integrals $\{I_1;\;I_1^\mu\}$
and $\{I_2;\;I_2^\mu;\}$ are connected with following relations
\begin{equation}
\frac{\partial I_1(k_1;\;M^2)}{\partial M^2}=I_2(k_1, k_1;\;M^2)
\label{dI_1}
\end{equation}
and
\begin{equation}
\frac{\partial I_1^\mu(k_1;\;M^2)}{\partial M^2}=I_2^\mu(k_1,
k_1;\;M^2). \label{dI_1mu}
\end{equation}
 These relations are
regularization independent and play a significant part in the
approach. It is easy to verify the validity of these relations for
traditional regularization schemes (four-dimensional cutoff,
Pauli-Villars regularization, etc) by straightforward calculation,
but their validity  is not evident  directly from above-mentioned
formulae of work \cite{BDK}. To prove these relations  we should
define derivatives of the external-momentum-independent integrals
 over $M^2$. For this purpose we note
that derivatives  of logarithmically-divergent integrals
 are  convergent integrals
and can be calculated without any regularization. The calculations
of these convergent integrals gives us zero value for derivatives
of tensorial logarithmically-divergent integrals. Then it is easy
to verify that the derivatives of  quadratically-divergent
integrals are the corresponding logarithmically-divergent
integrals. Taking into account these circumstances
 it is
easy to prove equations  (\ref{dI_1}) and (\ref{dI_1mu}) for BDK
expressions of integrals (\ref{I1}) and (\ref{I2}). The
significance of relations (\ref{dI_1}) and (\ref{dI_1mu}) are
evident since they can be used for alternative definitions of
quadratically divergent integrals $I_1$ and $I_1^\mu$ without
additional regularization.

With obtained expressions for integrals (\ref{I1}) and (\ref{I2})
the authors of work \cite{BDK} have analyzed the symmetry
constraints on one-loop Green functions and have argued the
necessity of
 consistency relations,  which are concluded in zero value for
 tensorial external-momentum-independent integrals
 (see eq. (86) in work \cite{BDK}).
With these relations all  symmetry properties of the theory (such
as Furry theorem, Ward identities, etc) are fulfilled for all
one-loop Green functions. This point is one of the important
results of work \cite{BDK}. From the point of view of above
discussion the BDK consistency relations  simply assert zero
values of corresponding integration constants.

The choice of parameters  of the NJL model with Lagrangian
(\ref{LSU2}) in the leading approximation is  principally defined
by two divergent integrals, namely, logarithmically-divergent
integral $I_2$  and quadratically-divergent integral $I_1$.
Integral  $I_1$ is a part of the gap equation
\begin{equation}
 M=m_0-8ign_cM\cdot I_1, \label{gap}
\end{equation}
which defines dynamical (constituent) quark mass $M$. Integral
$I_2$ determines the structure of meson propagators. Both these
integrals can easily  be defined with differential regularization
without using the above-mentioned algebraic identity for the
propagator function.

To define $I_2$ one can use well-known trick, which goes back to
 Gelfand and Shilov \cite{GeShi} (see also \cite{Collins}). Namely,
 let introduce new external variables
\begin{equation}
 p=k_1-k_2,\;\;P=k_1+k_2. \label{p}
\end{equation}
Derivatives  $\partial I_2/\partial p_\mu$ ¨ $\partial
I_2/\partial P_\mu$ are convergent integrals, and their
calculation gives us
\begin{equation}
 \frac{\partial I_2}{\partial p_\mu}=\frac{ip_\mu}{16\pi^2}\int_0^1
du \frac{u(1-u)}{M^2-u(1-u)p^2}, \;\;\; \frac{\partial
I_2}{\partial P_\mu}=0.\label{dI_2}
\end{equation}
Basing on these results and using identity $ p^2\frac{\partial
f}{\partial p^2}=\frac{1}{2}p_\mu\frac{\partial f}{\partial
p_\mu}, $ we naturally go to the following definition:
\begin{equation}
I_2(p^2)=-\frac{i}{16\pi^2}\int_0^1du\log
\frac{M^2-u(1-u)p^2}{M_0^2}, \label{I2diff}
\end{equation}
where $M_0$ is the integration constant. This expression coincide
with the BDK ones  if constant $M_0$ is related to the scalar
external-momentum-independent integral $I_{log}$ (see eq. (40) in
work \cite{BDK}) as
\begin{equation}
I_{log}=-\frac{i}{16\pi^2}\log \frac{M^2}{M_0^2}=I_2(0).
\label{Ilogdiff}
\end{equation}

An essential feature of the above calculation is the permutation
of two limits -- differentiation and regularization removing.
Sure, such permutation is implied also in work \cite{BDK} (in the
calculations of finite parts). This permutation is an essence of
Gelfand-Shilov differential regularization \cite{GeShi}, which is
based on the infinite differentiability of generalized functions.

To define  $I_1$ we use regularization-independent relation
(\ref{dI_1}), which gives (with taking into account eq.
(\ref{Ilogdiff}))
\begin{equation}
I_1=\frac{i}{16\pi^2}(M^2\log \frac{M^2_0}{M^2}+M^2-M_1^2),
\label{I1diff}
\end{equation}
where $M_1^2$ is the integration constant of eq. (\ref{dI_1}). By
taking into account the consistency relations  such definitions
also corresponds to BDK ones.

 To calculate $M_0$ and $M_1$
 and, therefore, to define $I_0$ and $I_1$ in full,
we can use in the chiral limit ($m_0=0$) two
regularization-independent relations of NJL model (see, e.g.
\cite{kle}), namely
\begin{equation}
f^2_\pi=\frac{4n_cM^2}{i}I_2(0) \label{fpi}
\end{equation}
where $f_\pi$ is the pion-decay constant, and
\begin{equation}
\chi=2c^3=\frac{8n_cM}{i}I_1 \label{chi}
\end{equation}
where $\chi=<0\vert\bar\psi\psi\vert 0>$  and $c$ is the quark
condensate.   We have from  equation (\ref{fpi})
\begin{equation}
I_2(p^2)=
\frac{i}{16\pi^2}\Bigl[\frac{4\pi^2f_\pi^2}{n_cM^2}-\int_0^1du\log(
1-u(1-u)\frac{p^2}{M^2})\Bigr]. \label{I0BDK}
\end{equation}
Then, using equations (\ref{I1diff}) and  (\ref{chi}), we obtain
for $M_1$
\begin{equation}
M_1^2=M^2+\frac{4\pi^2f_\pi^2}{n_c}-\frac{4\pi^2c^3}{n_cM}.
\label{BDKE}
\end{equation}
This equation (we shall refer it as BDK equation) plays a key role
in the BDK approach.\footnote{An alternative way to introduce the
implicit regularization can be based on an approach of work
\cite{Farhi}.}

 At pion constant $f_\pi$  and
quark condensate $c$ being fixed equation (\ref{BDKE}) is the
equation for dynamical quark mass $M$, where $M_1$ is a parameter.
In terms of new variable $ x=-\frac{(n_c/2\pi^2)^{1/3}}{c}\;M$
equation (\ref{BDKE}) can be written as
\begin{equation}
x^3-3ax+2=0, \label{BDKEr}
\end{equation}
where
$a=\frac{(n_c/2\pi^2)^{2/3}}{3c^2}\;(M_1^2-\frac{4\pi^2f_\pi^2}{n_c})$.
Depending on the value of parameter  $a$, three cases are possible:\\
 (1) at $a<1$ equation (\ref{BDKEr}) possesses
 one real negative root  $x_1<0$;\\
  (2) at $a=1$ equation (\ref{BDKEr}) possesses one negative root
   $x_1=-2$ and one positive root $x_2=1$.\\
(3) at  $a>1$ equation (\ref{BDKEr}) possesses one negative root
$x_1<0$  and two positive roots $0<x_{2}<1$ and $x_3>1$.

It is clear, that at $c<0$  the second case ($a=1$) only can be
physically accepted. The authors of work \cite{BDK} take only this
situation  as a solely possible choice for model parameters. The
value of the dynamical quark mass, which corresponds to positive
root $x_2=1$ at $ n_c=3$, is $ M=-1.873\;c$. At condensate value
 $c=-250\;MeV$ it gives  $M= 468\;MeV$. Coupling constant $g$ can
 be defined from the well-known relation of the NJL model
\begin{equation}
 g=-M/2c^3, \label{g}
\end{equation}
 which is also regularization-independent (see, e.g., \cite{kle}).
Noteworthy, the quark mass value and the coupling value  depend on
the quark condensate value only, and do not depend on the pion
constant value. The last one defines the value of $M_1$, which is
$M_1=879\;MeV$ at $f_\pi=93 \;MeV.$

It is interesting to compare the parameter values of the BDK
approach with the parameter values of traditional regularizations.
The value of quark mass (468 MeV) in the BDK approach is
noticeably higher in comparison with the values in the
four-dimensional momentum cutoff regularization (236 MeV) and the
Pauli-Villars regularization (240 MeV) (at given value of quark
condensate --250 MeV). Apart from the quark mass dependence on the
condensate value is quite different. In the BDK approach the quark
mass is proportional to the condensate, whereas in 4-momentum
cutoff and Pauli-Villars regularization  the quark mass increases
at decreasing the absolute value of condensate. For instance, at
condensate value $c=-210\; MeV$ the quark mass is $M=393\; MeV$ in
the BDK approach and $M=423\; MeV$ in the four-dimensional
momentum cutoff. At the same time, from the point of view of the
phenomenology the parameters' values in BDK approach are quite
reasonable. Apart from it is necessary to recognize that
expressions for the Green functions are much more simple in
comparison with any traditional regularization.

In the framework of traditional regularizations the
chiral-symmetry-breaking phase always co-exists with the
chiral-symmetric phase, which is energetically unfavored. The
existence of the symmetric phase is not evident for the implicit
regularization in the original formulation of work \cite{BDK}. But
the  transition to the symmetric phase with $M=0$ (in  chiral
limit $m_0=0$) can be easily performed with the proposed
differential formulation of the implicit regularization. Really,
taking $M=0$ in equation (\ref{dI_2}), we obtain $
 \frac{\partial I_2}{\partial p_\mu}=-\frac{ip_\mu}{16\pi^2p^2}
$ and, therefore,
$I_2(p^2)=-\frac{i}{8\pi^2}\log\frac{p^2}{M^2_0}$ instead of
(\ref{I2diff}). Then $I_1=-\frac{iM^2_1}{16\pi^2}$ at
$M\rightarrow 0$ and from equations (\ref{fpi}) and (\ref{chi}) it
follows that $f_\pi\rightarrow 0$ and $c\rightarrow 0$ in
correspondence with the trivial physics of the symmetric phase.

An essential difference from other regularizations manifests  the
properties of Green functions in the Euclidean momentum region. In
the BDK approach the meson propagators possess a non-physical
singularity -- a pole at the negative momentum square (Landau
pole, or Landau ghost).\footnote{The existence of such pole was
discovered firstly in quantum electrodynamics \cite{Landau} and
presents a characteristic feature of theories without  an
asymptotic freedom in deep-Euclidean region. The existence of the
Landau ghost in a system of fermions coupled to a chiral field has
been observed in work \cite{Ripka}} The meson propagators in the
NJL model are scalar and pseudoscalar parts of the two-particle
amplitude and can be written as
\begin{equation}
D_\sigma=-\frac{ig}{m_0/M-4ign_c(4M^2-p^2)I_2(p^2)} \label{sigma}
\end{equation}
for the sigma-meson and
\begin{equation}
 D_\pi=-\frac{ig}{m_0/M+4ign_cp^2I_2(p^2)}. \label{pi}
\end{equation}
for the  pion.  In the chiral limit the equation for the Landau
pole, as it follows from equation (\ref{I0BDK}), is
\begin{equation}
z\log\frac{z+1}{z-1}=2+\frac{1}{n_c}\Biggl( \frac{2\pi
f_\pi}{M}\Biggr)^2, \label{Landau}
\end{equation}
where $z=\sqrt{1-\frac{4M^2}{p^2}}$. At above values of the model
parameters ($n_c=3,\;f_\pi=93 \;MeV,\; M=468\; MeV$) this equation
has solution  $p_L^2=-4.29 M^2=-(969\; MeV)^2$, i.e., Landau mass
value $M^2_L=-p^2_L$ approximately twice larger in comparison with
the quark mass: $M_L=2.07M$. Beyond the chiral limit (at $m_0\neq
0$) this value changes very small (about 1\%) due to the smallness
of current quark mass $m_0$.

The existence of the Landau pole is a serious problem of the BDK
approach. In particular, any calculations with meson loops become
problematic. Though such calculations exceed the framework of the
one-loop approximation, which is a subject of work \cite{BDK}, the
impracticability  of these calculations means a  principal
impossibility of  calculations of corrections to leading
approximation and cannot be acceptable.

Further, the Landau pole in quantum electrodynamics due to the
smallness of fine structure constant $\alpha$ is located in the
very distant asymptotic Euclidean region (($M^2_L)^{QED}\simeq
-m^2_e\exp\{\frac{3\pi}{\alpha}\}$, where $m_e$ is the electron
mass and  $\alpha\simeq 1/137$), and its presence can be in
principle ignored. Really, at the  much smaller energies the
quantum electrodynamics becomes a part of an asymptotically free
grand unification theory with  self-consistent asymptotic
behavior. In the NJL model with the implicit regularization the
Landau pole is located near the physical region of the model, and
similar reasoning is impossible in principle. From the other hand,
the traditional regularizations, such as the four-dimensional
cutoff or the Pauli-Villars regularization, are free on this
problem -- the meson propagators in these regularizations do not
have the Landau poles. Therefore, implicit regularization,
possessing the certain appeal and simplicity, contains serious
defect as the nearby Landau pole.

To improve the situation a compromise approach is necessary, which
conserves main features of implicit regularization and at the same
time solves the problem of the Landau pole. Such compromise can be
achieved with the Feynman regularization for logarithmically
divergent integral $I_2$:
\begin{equation}
I_2(p^2)= \int \frac{d^4q}{(2\pi)^4}
\Big\{\frac{1}{(M^2-(p+q)^2)(M^2-q^2)}-
\frac{1}{(M^2_r-(p+q)^2)(M^2_r-q^2)}\Big\},
\label{I0bar}
\end{equation}
where $M_r$ is a regulator mass ($M^2_r>M^2$), and  the definition
of quadratically divergent integral  $I_1$  as before is made by
relation (\ref{dI_1}).

In Euclidean region $p^2<0$ integral
 (\ref{I0bar}) can be represented as
 $$I_2=\frac{i}{16\pi^2}F(-\frac{p^2}{M^2}),$$
where
\begin{equation}
F(x)=\int_0^1du\log\frac{M^2_r/M^2+u(1-u)x}{1+u(1-u)x}.
\label{F(x)}
\end{equation}
Prove the absence of zeroes of this function at  $x>0$. Evaluating
elementary integrals in equation  (\ref{F(x)}) and introducing
variables $v=\frac{1}{2}(\sqrt{1+\frac{4}{x}}-1)$ and
$v_r=\frac{1}{2}(\sqrt{1+\frac{4M^2_r}{xM^2}}-1)$ we obtain the
following expression:
\begin{equation}
F=2\Bigl[\log\frac{1+v_r}{1+v}+v_r\log(1+\frac{1}{v_r})-v\log(1+\frac{1}{v})\Bigr].
\end{equation}
If $M^2_r>M^2,$ then $v_r>v$, and from elementary inequalities
$\log\frac{1+v_r}{1+v}>0$ and
$v_r\log(1+\frac{1}{v_r})>v\log(1+\frac{1}{v})$ it follows that
 $F>0$ in the Euclidean region, i.e. the  meson propagators do not possess
 the Landau pole with this definition. For other aspects this modified
 regularization is similar to implicit regularization.
 For the modified
 regularization
\begin{equation}
I_2(0)=-\frac{i}{16\pi^2}\log \frac{M^2}{M_r^2}, \label{I_2(0)}
\end{equation}
and, consequently,
 $I_1$ is defined by same formula
 (\ref{I1diff}) with substitution
$$M_0\rightarrow M_r,$$ i.e. integration constant  $M_0$
everywhere is substituted by regulator mass
 $M_r$. Such re-definition of $I_1$ still implies an existence of
 additional parameter $M_1^2$, which is the integration constant
 of equation ({\ref{dI_1}).

Equation  (\ref{BDKE}), which determines quark mass $M$, has
exactly the same form for this modified regularization, and,
consequently, all parameter values are the same. Remark, condition
 $M^2_r>M^2$ is the automatical consequence of formula (\ref{fpi}).
 Therefore, the proposed modification conserves main features of
 the implicit regularization and simultaneously solves the
 problem of Landau
 pole.\footnote{Another way to remove the Landau ghost in the
 NJL model is the method
 of work \cite{Bog}, based on the Kallen--Lehmann representation.
 This method has been applied in work \cite{Hartmann} to the
 chiral $\sigma$-model. However, the proposed receipt seems to be
 preferable in the framework of the implicit regularization since
 this regularization has deal with a set of integrals while the
 method of works \cite{Bog, Hartmann} should be applied to the
 proper two-point functions.}

A general receipt for definitions of divergent integrals
(\ref{I1}) and (\ref{I2}) can be formulated as follows: integrals
(\ref{I2}) are defined by formulae
\begin{equation}
\{I_2;\;I_2^\mu;\;I_2^{\mu\nu}\}=
\{I_2;\;I_2^\mu;\;I_2^{\mu\nu}\}^{BDK}(M^2)-
\{I_2;\;I_2^\mu;\;I_2^{\mu\nu}\}^{BDK}(M^2_r). \label{RecI2}
\end{equation}
Upper index $BDK$ means that each integral with mass $M$ or $M_r$
is defined by formulae of work \cite{BDK}.\footnote{With such
definitions integrals $I_2$ and $I_2^\mu$ are convergent without
additional regularization. Since for self-consistency of the
receipt it is necessary to verify the coincidence of usual
calculation of these integrals with the calculation by formula
(\ref{RecI2}), based on the BDK definition of divergent integrals.
Simple calculation demonstrates the reality of such coincidence.}

Then integrals (\ref{I1}) are defined by equations (\ref{dI_1})
and  (\ref{dI_1mu}). A consistency relation is the coincidence of
integration constants, and, therefore,
\begin{equation}
I_1^\mu=-k_1^\mu I_1 \label{RecI1}
\end{equation}
as in the BDK approach with consistency conditions.

It should be noted, that due to linearity of
 consistency constraints considered in work \cite{BDK},
the analysis of  symmetry conservation,  which was made in this
work, can be carried to the proposed modification.

A principal distinction between the implicit regularization and
the proposed ghostless modification consists in the treating of
integral $I_2$. Parameter $M_0$ in the implicit regularization is
the integration constant whereas parameter $M_r$ in the Feynman
regularization (\ref{I0bar}) is introduced "by hands" and seems to
be an outside object. The question arises, if really this object
is some "secret parameter", which  ensures the self-consistency of
the NJL model on the quantum level? We cannot answer to this
question now with the  full confidence, but the structure of the
implicit regularization gives us an allusion to the possible
realization of this idea. The term with regulator mass $M_r$ in
the definition of integral $I_2$ by formula (\ref{I0bar}) can be
interpreted as the rise of some repulsive interaction (i.e.
interaction with a negative coupling) at the one-loop level. The
coupling constant of the NJL model is connected with the quark
mass and the condensate by formula (\ref{g}). If the quark
condensate is negative the negative coupling corresponds to the
negative quark mass. In this connection it is not out of place to
recall about the negative solution of BDK equation (\ref{BDKE})
("quark ghost"). In the physical  case this solution is $-2M$,
where $M$ is the dynamical quark mass. The existence of this
solution can be an indication of such repulsive interaction, which
becomes apparent on the one-loop quantum level. At given above
parameter values we have $|M_r|\simeq 1.3 M$. This value, of
course, is far from the expected value. When the absolute value of
the quark condensate decreases  ratio $|M_r|/M$ arises, and at
$c=-210 \;MeV$ it equals to $\simeq 1.5$. At  $c=-150\; MeV$ it
reaches expected value 2, but the last value of the condensate is
phenomenologically unacceptable since it leads to the very large
value of the current quark mass $m_0 = 21\; MeV$. Therefore,
  the question of the possible  "inter-termination of ghosts"
  \quad in the framework of the NJL
model remains to be open.

In conclusion,  discuss  briefly the definition of model
parameters beyond the chiral limit. At $m_0\neq 0$ from
eq.(\ref{pi}) it follows the pion spectrum equation
\begin{equation}
m_0=\frac{2in_cM^2}{c^3}\cdot m^2_\pi \cdot I_2( m^2_\pi).
\end{equation}
Practically this equation is used for definition of current quark
mass  $m_0$.

Taking approximation $I_2(m_\pi^2)\simeq I_2(0)$, we obtain the
well-known current-algebraic Gell-Mann-Oakes-Renner formula
\begin{equation}
m_\pi^2f^2_{\pi}\simeq -m_0<\bar{\psi}\psi>.\label{GMOR}
\end{equation}
This relation, of course, is regularization-independent (and what
is more, model-independent). At $m_\pi=135\;MeV,\;c=-250 \;MeV$ we
obtain from eq.(\ref{GMOR}) $m_0=5.2\;MeV$. The exploiting of
regularization-dependent exact value $I_0(m_\pi^2)$ changes this
result very slightly (approximately of 1 percent) due to the
smallness of pion mass, and such specification, of course, lays
out the framework of the model, it being right as for the implicit
regularization as the modified variant.

The sigma-meson mass in the chiral limit is $2M$. Beyond the
chiral limit the equation on the sigma-meson mass follows from
equation (\ref{sigma}):
\begin{equation}
m_0=\frac{2in_cM^2}{c^3}\cdot (m^2_\sigma-4M^2) \cdot I_2(
m^2_\sigma).
\end{equation}
With rather crude approximation  $I_2( m^2_\sigma)\simeq I_2(0)$
and with taking into account equation (\ref{GMOR}) we obtain the
well-known
 regularization-independent formula (see, e.g.,
 \cite{kle})
\begin{equation}
m^2_\sigma\simeq 4M^2+m^2_\pi.
\end{equation}
The much more exact (but regularization-dependent) approximation
is
 $I_2( m^2_\sigma)\simeq I_2(4M^2)$. In this case, we obtain
\begin{equation}
m^2_\sigma\simeq 4M^2+0.21m^2_\pi
\end{equation}
 for
the  implicit regularization, and
\begin{equation}
m^2_\sigma\simeq 4M^2+0.26m^2_\pi
\end{equation}
 for the  modified variant.

As we see from the above formulae,  in the  physical region of
energies about 1 GeV the quantitative difference of these
regularizations is not very significant.

\section*{Acknowledgments}

Author is grateful to R~Jafarov and V~A~Petrov for useful
discussions.


\begin{thebibliography}{99}

\bibitem{BDK} Battistel O A, Dallabona G and Krein G 2008
 {\it Phys.Rev.} D {\bf 77} 065025
(hep-th/0803.0537)
\bibitem{kle} Klevansky S P 1992 {\it Rev.Mod.Phys.} {\bf 64} 649
\bibitem{BatNem} Battistel O A and Nemes M C 1999
{\it Phys.Rev.} D {\bf 59} 055010;\\
 Farias R L S, Dallabona G, Krein
 G and Battistel  O A  2006
{\it Phys.Rev.} C {\bf 77} 065201
\bibitem{GeShi} Gelfand I M and Shilov G E  1964 {\it Generalized
Functions} (Academic, San Diego)  Vol. 1
\bibitem{Collins} Collins J C 1984 {\it Renormalization} (Cambridge Univ.
Press)
\bibitem{Farhi} Farhi E, Graham N, Jaffe R L and Weigel H
 2002 {\it Nucl.Phys.} B {\bf 630} 241
\bibitem{Landau} Landau L D and Pomeranchuk I Ya 1955
{\it Dokl.Akad.Nauk Ser.Fiz.} {\bf 102}  489
\bibitem{Ripka} Ripka G and Kahana S 1987 {\it Phys.Rev.} D {\bf
36} 1233
\bibitem{Bog} Redmond P J 1958 {\it Phys.Rev.} {\bf 1958} 1404;\\
 Bogoliubov N N, Logunov A A and Shirkov D V 1959
{\it ZhETF} {\bf 37} 805
\bibitem{Hartmann} Hartmann J, Beck F and Bentz W 1994 {\it
Phys.Rev} C {\bf 50} 3088


\end{thebibliography}
\end{document}